# LLM-Generated Samples for Android Malware Detection

Nik Rollinson and Nikolaos Polatidis*

School of Architecture, Technology and Engineering

University of Brighton, U.K

*Corresponding author: n.polatidis@brighton.ac.uk

## Abstract

Android malware continues to evolve through obfuscation and polymorphism, posing challenges for both signature-based defenses and machine learning models trained on limited and imbalanced datasets. Synthetic data has been proposed as a remedy for scarcity, yet the role of large language models (LLMs) in generating effective malware data for detection tasks remains underexplored. In this study, we fine-tune GPT-4.1-mini to produce structured records for three malware families: BankBot, Locker/SLocker, and Airpush/StopSMS, using the KronoDroid dataset. After addressing generation inconsistencies with prompt engineering and post-processing, we evaluate multiple classifiers under three settings: training with real data only, real-plus-synthetic data, and synthetic data alone. Results show that real-only training achieves near perfect detection, while augmentation with synthetic data preserves high performance with only minor degradations. In contrast, synthetic-only training produces mixed outcomes, with effectiveness varying across malware families and fine-tuning strategies. These findings suggest that LLM-generated malware can enhance scarce datasets without compromising detection accuracy, but remains insufficient as a standalone training source.

## 1. Introduction

Android malware continues to outpace signature-based defences, with families evolving through obfuscation, polymorphism, and rapid varianting. Although Machine Learning (ML) and Deep Learning (DL) pipelines have advanced detection accuracy on benchmark datasets, their effectiveness depends heavily on timely, high-quality labels and balanced class distributions, conditions rarely achieved in operational settings. Data scarcity is particularly acute for specific malware families and behaviours, limiting classifier robustness and hindering reproducibility across studies.

Recent work has explored synthetic intrusion data and adversarial generation to address these challenges, yet there remains no empirical evaluation of how synthetically generated Android malware, produced specifically by Large Language Models (LLMs), influences detection performance. This research addresses that gap by generating structured malware records with a fine-tuned LLM and systematically assessing their utility in training and augmenting ML classifiers. Using the KronoDroid dataset for its hybrid static–dynamic feature space and balanced malicious/benign splits, we fine-tune GPT-4.1-mini to emulate the characteristics of three families, BankBot, Locker/SLocker, and Airpush/StopSMS, resolve generation pathologies through prompt engineering and post-hoc filtering, and evaluate the downstream impact across five classifiers under three scenarios: real-only training, real-plus-synthetic augmentation, and synthetic-only generalisation.

Through rigorous experimentation, this paper provides the following contributions:

- Augmenting real samples with LLM-generated malware to enhance performance.
- Train machine learning algorithms on synthetic malware and detect malware effectively.

The rest of the paper is structured as follows: Section 2 contains the related works, section 3 delivers the methodology, section 4 presents the experimental results and section 5 contains the conclusions and future work.

## 2. Related Work

Artificial intelligence (AI) has emerged as a transformative force in the field of cybersecurity, enabling new methods for detecting, preventing, and even simulating cyber threats (Achuthan *et al.*, 2024). This review explores the intersection of AI and cybersecurity across several key areas. Section 2.1 discusses foundational approaches using Machine Learning (ML), Deep Learning (DL), and Large Language Models (LLMs) for threat detection. Section 2.2 delves into the potential misuse of LLMs in malware generation and automation. Section 2.3 examines AI-based techniques for malware detection and prevention. Section 2.4 focuses on the use of synthetic data in cybersecurity. Finally, Section 2.5 identifies a key research gap related to the use of LLM-generated synthetic malware for training detection models.

### 2.1 AI and Cybersecurity Foundations

With the increasing popularity of Large Language Models, the past few years saw a growing number of applications for them that can be employed in both defensive and offensive strategies (Mudassar Yamin et al., 2024). With the escalating sophistication and frequency of cybersecurity threats, conventional defence mechanisms that depend predominantly on manual analysis are proving progressively insufficient. Artificial intelligence technologies present promising advancements by enabling more resilient threat detection through intelligent data analysis, facilitating pattern recognition and the anticipation of future attacks, thereby mitigating the inherent constraints of traditional approaches (Sarker, 2023). As some attacks are highly complex, their manual detection that requires monitoring many security alerts and their analysis may be slow (Ali, Shah and ElAffendi, 2025). Conventional approaches frequently overlook attacks that deviate from established signatures, whereas AI techniques possess the capacity to generalise from prior examples (Ferrag et al., 2025). AI is also able to detect subtle anomalies that might be challenging to notice manually. Furthermore, among other useful functionalities the authors specifically highlight the capability of LLMs in generating synthetic data (Ankalaki et al., 2025).

## 2.2 LLMs for Malware Generation and Automation

Botacin (2023) evaluated OpenAI's GPT-3 model for malware generation, finding that it struggles to produce fully functional malware from broad descriptions but succeeds when prompts are decomposed into specific tasks. He asserts that existing models are incapable of generating malware from generic prompts but can alter existing malware to create novel variants. This raises concerns about the increasing accessibility of AI tools for malware development (Botacin, 2023). Ubavić et al. (2023) provide an example where a hacking community user created functional data-stealing malware using ChatGPT, automating a task that would otherwise require programming skills. However, the process still required iterative prompting. (Ubavić *et al.*, 2023). On the other hand, Pa et al. (2023) evaluated several models and tools, including Auto-GPT, an autonomous agent built upon OpenAI's GPT-3.5 and GPT-4, which can generate prompts and decompose broad tasks into manageable subtasks. The authors demonstrated that Auto-GPT can create malware from a generic prompt (Pa et al., 2023). Another capability of LLMs is to alter existing malware to generate novel variants. A practical implementation was demonstrated by CyberArk where the researchers successfully produced polymorphic malware by repeatedly instructing ChatGPT to mutate code, thereby generating new variants with each iteration and incorporating additional constraints to circumvent signature-based detection (Hilario *et al.*, 2024).

## 2.3 AI-Based Malware Detection and Prevention

Berrios et al. (2025) conducted a thorough literature review on AI-based malware detection approaches, covering deep learning models, generative adversarial networks, behavioural and anomaly-based detection, large language models, and their combinations. They note that the field is advancing with improving detection rates, and describe two principal approaches: static analysis, which involves the examination of a file's features, and dynamic analysis, which monitors events and behaviours within a controlled environment (Berrios *et al.*, 2025). Shaukat, Luo and Varadharajan (2023) report the successful integration of static and dynamic malware analysis methods by employing convolutional neural networks for feature extraction followed by support vector machines for classification, achieving a 16.5% improvement in accuracy on the Malimg dataset compared to conventional techniques (Shaukat, Luo and Varadharajan, 2023). Gyamfi et al. (2023) state that machine learning-based detection systems can analyse vast data volumes to identify malicious patterns, including zero-day vulnerabilities and polymorphic malware, offering rapid and adaptive detection that addresses limitations of traditional signature-based methods. However, they highlight challenges such as limited or imbalanced datasets and poor generalisation to novel or manipulated malware, leading to models performing well on benchmarks but struggling in real-world conditions, underscoring the need for greater robustness (Gyamfi *et al.*, 2023). Salem et al. (2024). Literature in this field agrees that AI methods, from Machine Learning to Deep Neural Networks, effectively detect and prevent malware but must be regularly updated with quality data to counter evolving attacks (Salem *et al.*, 2024).

## 2.4 Synthetic Data in Cybersecurity

Research conducted by Chalé and Bastian generated synthetic network flow data using CTGAN and TVAE for intrusion. The researchers found that ML classifiers that were trained both on real and synthetic data had the same performance as classifiers trained only on the real data. However, they also identified that purely synthetic training underperforms unless at least 15% of real data is retained. They also state that synthetic data does improve statistic learning, but it does not introduce new predictive information (Chalé and Bastian, 2022). Ammara, Ding and Tutschku (2024) reinforce these points in their comparative analysis of synthetic data generation in the cybersecurity field. They caution that similarity scores or synthetic-only results cannot substitute for real-world validation, and agreeing that synthetic data is useful for augmenting but not replacing real data (Ammara, Ding and Tutschku, 2024).

In contrast, Rahman et al. (2024) conducted research on using only synthetic data for Network Intrusion Detection Systems. GAN-based synthetic data was then used to train classifiers, and their performance was assessed against datasets with real data. They were able to achieve high accuracy rates, proving the possibility of detecting intrusion with classifiers trained only on synthetic data. This directly contradicts findings which concluded that purely synthetic training was not viable (Rahman et al., 2024). Almorjan, Basheri and Almasre, (2025) conducted research on using LLMs to create synthetic data of social media interactions that could help in identifying Indicators of Compromise (IoC), where they were looking for messages in social media chats that contain IP addresses or other information that could indicate a threat. The research used OpenAI's GPT 3.5 model with fine-tuning. The study resulted in two synthetic datasets, and the best accuracy of identifying the correct classes of IoC was 77% for the first dataset and 82% for the second. However, unlike studies evaluating against real data, they did not assess generalisation to real-world scenarios. The study suggests that purely synthetic data may be sufficient, however, the authors do not provide an external validation for that (Almorjan, Basheri and Almasre, 2025).

## 2.5 Research Gap

The literature review indicates that there is a gap in available research. We could not find any works that use Large Language Models to create synthetic malware records and use them to determine how they affect accuracy and other metrics of malware detection.

# 3 Methodology

## 3.1 Dataset Selection

As we have discussed above the quality of the original data is crucial (Rahman et al., 2024). Therefore, it is essential to analyse and select the right dataset with Android malware samples to use for further synthetic data generation and accuracy evaluations.

Elnashar, White and Schmidt (2025) examined prompt styles and output formats for GPT-4o with different file formats such as JSON, YAML and Hybrid CSV/Prefix. While many use JSON and YAML because they are structured and readable, the research investigated Hybrid CSV/Prefix as a less resource-intensive alternative. Hybrid CSV/Prefix format combines CSV and prefixed identifiers. They found that CSV/Prefix, the format with fixed schema and row-per-sample organisation delivers the best results for synthetic data generation tasks with large language models and is the most efficient in terms of token usage and processing time (Elnashar, White and Schmidt, 2025). Based on this, we should prioritise datasets that offer row-per-sample organisation.

Because Shaukat, Luo and Varadharajan (2023) showed improvement of detection rate with hybrid approach, we should consider prioritising a dataset that has both static and dynamic features. Furthermore, even though this research aims to be a proof of concept, we should have scalability in mind, and therefore, should aim at a well-balanced dataset so that the application discussed in this research could be applied to the whole dataset. And finally, dataset features should be human-interpretable to provide a better understanding of underlying data.

The Drebin dataset is often considered the baseline for Android malware datasets. However, it is not well-balanced with only 5,560 samples being malware from the total of 123,453 apps analysed, and uses only static features (Arp et al., 2014). MalGenome contributed early behavioural insights but is malware-only containing 1,260 malwares from 49 families (Zhou and Jiang, 2012). AndroZoo provides the largest number of analysed APKs, making it ideal as a source repository, however, it is not a feature table ready for use out of the box, and it lacks uniform labels and dynamic traces (Alecci et al., 2024).

AndroDex explicitly targets obfuscation, releasing 24,746 samples by converting DEX bytecode into images. Although it is robust for some use cases, there is no tabular feature matrix (Aurangzeb et al., 2024).

Two newer datasets explicitly target reproducible, human-readable features. MaDroid offers 50,429 labelled system-call sequences collected over 14 years across 10 Android marketplaces and is maliciousness-aware using VirusTotal rating as an attribute. It has a healthy class balance with 24,789 benign and 25,640 malicious samples. Features for each event span over several rows, they are human-interpretable to analysts but are not a compact table (Duan et al., 2024). Another example is KronoDroid. It combines per-sample timestamps with a fixed hybrid table of 489 columns combining static and aggregated dynamic features. Splits are well-balanced: the emulator dataset contains 28,745 malware and 35,246 benign samples, while the real-device dataset contains 41,382 malware and 36,755 benign samples. Many features are human-readable and systematically structured within CSV files (Guerra-Manzanares, Bahsi and Nõmm, 2021). Based on our requirements, KronoDroid is the most suitable dataset for our research.

### 3.2 Data Preparation

The dataset encompasses static and dynamic analysis features extracted from malware samples executed on virtual and physical Android devices, providing a realistic representation of behavioural artefacts such as system calls, permissions, and network interactions (Guerra-Manzanares, Bahsi and Nõmm, 2021). Specifically, we focused on the real device malware subset, which includes detailed feature vectors from infected applications, and the benign samples subset, comprising legitimate Android applications. The benign subset totals 36,755 samples, from which random selections (without fixed random state, via Python scripting) were drawn to maintain experimental balance.

To ensure behavioural diversity, we selected three distinct malware families from the KronoDroid dataset: BankBot (1,297 samples, active primarily from 2013 to 2020), Locker/SLocker (1,846 samples, spanning 2008 to 2020, with some anomalous timestamps potentially indicating extraction errors, e.g., pre-2008 dates), and Airpush/StopSMS (7,775 samples, active primarily from 2008 until 2016, with a variety of anomalous timestamps present in the dataset). BankBot, an Android banking trojan, primarily employs overlay attacks to intercept user credentials on financial applications, often tries to appear as benign software with delayed activation mechanisms (Bai et al., 2021). In contrast, Locker/SLocker represents an early form of Android ransomware, characterised by device UI locking via full-screen overlays, activity hijacking, key disabling, and exploitation of device administrator privileges, including file encryption of media and documents on external storage (Su et al., 2019). Airpush, on the other hand, is an advertising library that obfuscates its ad-library packages to avoid identification by package names, while SMS trojans disguise themselves as legitimate apps and secretly send SMS messages without user consent, sometimes also reading and making phone calls (Rastogi *et al.*). Families were treated separately to assess detection performance across different malware types.

A 1:1 ratio between malware and benign samples was maintained across configurations to mitigate class imbalance. Benign samples were selected randomly from the entire subset, without temporal filtering, as date-related features were excluded from analysis.

### 3.3 Synthetic Data Generation Using Large Language Models

Synthetic malware samples were generated using OpenAI's GPT-4.1-mini model (base version dated 2025-04-14) to augment the merged real malware set, addressing data scarcity in specialised families while preserving privacy and ethical constraints on real sample distribution. Direct attempts to emulate tabular structures via CSV input and output yielded highly inconsistent results, such as erroneous column counts ranging from 179 to over 7,000, necessitating a conversion to JSON format for improved structural fidelity. This aligns with prior research highlighting limitations of LLMs in handling non-sequential, tabular data formats due to their text-based training paradigms (Elnashar, White and Schmidt, 2025).

Fine-tuning on JSONL data failed OpenAI's moderation due to policy violations inferred from terms like 'malware', which triggered safety filters. To circumvent this, feature names were sanitised with neutral equivalents: for BankBot, 'malware' to 'app', 'Malware' to 'AppType', 'BankBot' to 'FinTech', 'MalFamily' to 'AppFamily', 'kill' to 'stop', and 'ptrace' to 'trace'; for Locker/SLocker, 'Locker/SLocker Ransomware' to 'HiddenTech' in place of 'FinTech'. This enabled moderation passage. Cost considerations estimated at $74.21 for the full 1,297 BankBot samples led to subsampling 50 representative records per family (total 100) for fine-tuning and using one epoch. To test whether larger finetuning sets and deeper training improve synthetic fidelity, we increased the finetuning set to 150 samples and ran 3 epochs for the Airpush/StopSMS family, keeping other settings constant. This creates an intentional, controlled departure from the BankBot and Locker setup (50 samples, 1 epoch) to study method sensitivity (family size, sample count, epochs) on downstream generalisation.

Despite successful training, the fine-tuned model exhibited instability, generating extraneous 'SYS_n' columns. These were mitigated by converting the data into the JSONL format. While that solved the original problem, the model returned repetitive outputs with identical data. This was then resolved by incorporating an exemplar record in the prompt and iterative refinement using OpenAI's prompt optimisation tool, which provided feedback on clarity and adherence without CLI invocation. Fine-tuning examples followed a structured message format as shown in figure 1 below.

```json
{
    "messages": [
        {
            "role": "system",
            "content": f"""You are a data-generation engine for Android application analysis records.
Output JSON with exactly {len(sanitised_cols)} keys. Keep AppType=1. Output only valid JSON."""
        },
        {
            "role": "user",
            "content": "Generate 1 Android FinTech app analysis record."  # Or 'HiddenTech' for Locker/SLocker
        },
        {
            "role": "assistant",
            "content": json.dumps([record])
        }
    ]
}
```

*Figure 1. Fine-Tuning Message*

Generation prompts were configured with temperature=0.7 for variety and max_tokens=16384. The system prompt enforced structure and realism as shown in figures 2 and 3.

```
You are a synthetic data generator for Android [FinTech/HiddenTech] application security analysis.

OUTPUT REQUIREMENTS:
- Return valid JSON object only (no markdown, no explanations)
- Use compact JSON format (no extra whitespace)
- Include ALL keys from the reference schema below
- All numeric values must be integers (no decimals)
- String values must be properly quoted
- AppType must always be 1

SCHEMA REFERENCE (for structure only - DO NOT copy these values):
{json.dumps(EXAMPLE_RECORD, separators=(',', ':'))}

GENERATION RULES:
- Generate completely unique synthetic values
- System call counts should reflect realistic Android [FinTech/HiddenTech] app behavior
- Permission counts should be consistent (nr_permissions = sum of permission grants)
- Detection_Ratio should be between 0.0 and 1.0
- Package name should follow Android naming convention (com.company.app)
- SHA256 should be 64-character hex string
- File sizes should be realistic for mobile apps (100KB - 50MB range)
- Dates should use MM/DD/YYYY format
- AppFamily must be "[FinTech/HiddenTech]"
- No null values - use 0 for unused numeric fields
```

*Figure 2. Generation Prompt*

```
user_prompt = f"Generate 1 unique Android [FinTech/HiddenTech] security analysis record #{record_num}.
Create realistic synthetic data that differs from the reference schema."
```

*Figure 3. User Prompt*

This yielded 392 synthetic records for BankBot and 301 for Locker/SLocker after removal of 12 duplicates. Data quality issues included repeated sha256 hashes (non-unique, with unrealistic patterns like multiple zeros) and occasional omissions in the AppType label (expected as 1 for all synthetic samples). Duplicates were minimal, and as labels are excluded from training, these did not impact detection evaluations.

To address potential LLM hallucinations (e.g., unrealistic syscall patterns), synthetic data was visually inspected by the author for qualitative similarity to real distributions (e.g., feature ranges and correlations), confirming approximate mimicry without formal statistical tests such as Kolmogorov-Smirnov or principal component analysis. Prompts explicitly instructed uniqueness, labelling, and absence of nulls, but adherence varied, underscoring LLM limitations for tabular synthesis.

### 3.4 Experimental Setup and Evaluation

After the synthetic data generation and model training, the dataset underwent processing to ensure consistency and suitability for machine learning. From the original 484 features, we excluded identifiers and metadata that could induce overfitting or biases: Malware (binary label), Detection_Ratio (antivirus detection percentage), MalFamily (family identifier), Scanners (detecting antivirus list), TimesSubmitted (submission count), NrContactedIps (network contacts), Package (application package name), sha256 (hash identifier), EarliestModDate, and HighestModDate (modification timestamps). This reduced the feature set to 474 numeric columns.

Non-numeric artefacts were addressed: 'None' values in columns such as Activities, NrIntServices, NrIntServicesActions, NrIntActivities, NrIntActivitiesActions, NrIntReceivers, NrIntReceiversActions, TotalIntentFilters, and NrServices, intended as counts of Android component invocations, were imputed to 0. We then filtered out all columns in which more than 70% of the values were zeros. This resulted in dropping the additional 87 columns, reducing the feature set to 387 feature columns. The dataset detailes are shown in table 1 below.

| Feature Category | Description | Exact Count |
| --- | --- | --- |
| **System Call Features** | System call usage indicating app behaviour at OS level. | 83 |
| **SYS_XXX & Misc Syscalls** | Obscure or custom syscall variants (SYS_300–SYS_369 and others). | 71 |
| **Android Permissions** | Declared Android permissions (normal, dangerous, signature, etc.). | 166 |
| **Permission Summary Metrics** | Aggregated counts and summaries of permission usage. | 7 |
| **App Structure & Manifest Features** | Manifest and structural metadata from the APK file. | 12 |
| **Other / Unclassified** | Unclassified or uncommon features not in main categories. | 48 |
| **Total** | | **387** |

Table 1. Resulting Dataset

Outliers were retained without removal, as elevated counts in features, such as system calls, likely reflect legitimate behavioural variations in Android applications rather than errors. Feature scaling was applied using StandardScaler from scikit-learn to normalise numeric values, addressing the wide variance in scales.

We evaluate three scenarios per family (BankBot, Locker/SLocker, Airpush/StopSMS) across 5 classifiers: 3 scenarios × 5 algorithms × 3 families = 45 total combinations, to assess synthetic data's impact on detection accuracy.

1. **Real malware vs. benign:** Real malware samples (label 1) paired with an equal number of random benign samples (label 0) for 1:1 balance. Stratified 80/20 train/test split on the combined balanced set.
2. **Real + synthetic malware vs. benign:** Concatenation of real and synthetic malware (both label 1) balanced 1:1 against benign (label 0). Stratified 80/20 train/test split on the combined balanced set.
3. **Train on synthetic, validate/test on real (synthetic → real generalisation):** Training on synthetic malware; real malware split 50/50 into validation and test sets (disjoint). Benign split 40/30/30 into train/validation/test (disjoint). Each split undersampled benign to match malware count for 1:1 balance.

**Comparability note:** Because Airpush/StopSMS used 150 finetuning samples and 3 epochs, synthetic to real results for Airpush are not directly comparable to BankBot/Locker synthetic to real under a 50-sample, 1-epoch regime; this contrast is by design to assess sensitivity to finetuning depth and data size.

Finally, five classifiers were tested: k-Nearest Neighbours (KNN), Decision Trees, Logistic Regression, Multilayer Perceptron (MLP), and Random Forest, all from scikit-learn. Models were wrapped in a pipeline: StandardScaler → Classifier. Hyperparameter tuning used grid search with 5-fold StratifiedKFold cross-validation (shuffle enabled) on the training set only, selecting by accuracy. The best estimator was re-fit on the full training set. Data-leakage prevention included hashing feature rows (pandas.util.hash_pandas_object) to verify no train/test intersection in 80/20 scenarios, and explicit disjoint subsets in synthetic to real. On the held-out test set, we reported accuracy, ROC AUC (from predict_proba), precision, recall, F1, and false positive rate (from confusion matrix). Nonparametric 95% confidence intervals (CIs) for test accuracy were computed via bootstrap resampling (B=1000). Confusion matrices were saved for qualitative analysis. In synthetic to real, metrics were reported for validation and test, with test as the final basis. Splits were stratified to preserve 50/50 balance; randomisation reflected script defaults with no random state used. Design choices such as balanced evaluation, uniform preprocessing, isolation, cross-validation, and CLs ensured robust assessment.

## 4 Experimental Evaluation

Experiments were run on an Apple M3 Pro processor with 18 GB RAM using macOS Sonoma 14.3. Python programming language has been used with pandas, numpy, sklearn, matplotlib, and seaborn libraries. Public version of the KronoDroid dataset was used for the experimentation[1].

The following sections demonstrate the detailed breakdown of experimental evaluation across five classifiers. Each of the individual classifier sections describe the settings used and outcomes of the evaluations. Later in the chapter we discuss the results.

### 4.1 Evaluation Metrics

For this research we used the Accuracy, ROC AUC, Precision, Recall, F1 Score, False Positive Rate, and Confidence Intervals metrics as shown by the equations 1, 2, 3, 4, 5, 6 and 7 below. Moreover, TP stands for True Positives, TN for True Negatives, FP for False Positives and FN for False Negatives.

(1) $Accuracy = (TP + TN)/(TP + TN + FP + FN)$
(2) $ROC\ AUC = \int_0^1 TPR(FPR^{-1}(t))\ dt$
(3) $Precision = TP/(TP + FP)$
(4) $Recall = TP / (TP + FN)$
(5) $F1\ Score = 2 \times (Precision \times Recall) / (Precision + Recall)$
(6) $FPR = FP / (FP + TN)$
(7) $CI_{95\%} = \left[\theta^*_{(2.5)}, \theta^*_{(97.5)}\right]$

### 4.2 k-Nearest Neighbors (kNN)

The K-Nearest Neighbour (KNN) algorithm is a supervised learning approach that classifies a new instance by comparing it to already labelled examples and assigning the class most common among its k closest neighbours. In practice, an application is converted into a feature vector, such as permissions, API calls, or intents, and its similarity to other applications is calculated, often with Euclidean distance. The decision is then based on the majority label of those nearest points (Babbar *et al.*, 2023). Its strength lies in the fact that it does not rely on prior assumptions about data distribution, allowing it to capture subtle patterns that might emerge from evolving malware families (Ucci, Aniello and Baldoni, 2019).

Experimental results confirm the usefulness of KNN in malware detection, outperforming several other classifiers (Babbar *et al.*, 2023). This combination of simplicity, interpretability, and accuracy highlights why KNN is a valuable tool in malware analysis pipelines. The detailed performance metrics for kNN are provided in the tables 2, 3, and 4 below, highlighting key indicators across scenarios. Each table represents results per each individual malware family. The Real vs Benign column represents the baseline scenario where both malware and benign samples were taken from the Kron-

---
[1] https://github.com/aleguma/kronodroid

oDroid dataset without synthetic augmentations. The Real + Synthetic vs Benign column represents results where the malware samples from the KronoDroid dataset were enriched with the synthetic data we generated. The Trained-on Synthetic Tested on Real vs Benign column represents the results where the model was trained exclusively using synthetic malware samples we generated and real benign samples and then tested on real malware and benign samples from the KronoDroid dataset. A 95% confidence interval (CI) is a range of values calculated from sample data that, in the long run, would contain the true population parameter in about 95% of repeated samples.

| Metric | Real vs Benign | Real + Synthetic vs Benign | Trained on Synthetic Tested on Real vs Benign |
|---|---|---|---|
| Accuracy | 0.9923 | 0.9911 | 0.7357 |
| ROC AUC | 0.9941 | 0.9985 | 0.7906 |
| Precision | 0.9923 | 1.0000 | 0.9811 |
| Recall | 0.9923 | 0.9821 | 0.4807 |
| F1 Score | 0.9923 | 0.9910 | 0.6453 |
| False Positive Rate | 0.0077 | 0.0000 | 0.0092 |
| 95% CI | [0.9903, 1.0000] | [0.9836 0.9970] | [0.7134 0.7589] |

Table 2. kNN Performance BankBot

| Metric | Real vs Benign | Real + Synthetic vs Benign | Trained on Synthetic Tested on Real vs Benign |
|---|---|---|---|
| Accuracy | 0.9715 | 0.9701 | 0.5119 |
| ROC AUC | 0.9862 | 0.9789 | 0.5265 |
| Precision | 0.9742 | 0.9609 | 0.6410 |
| Recall | 0.9687 | 0.9800 | 0.0542 |
| F1 Score | 0.9714 | 0.9704 | 0.0999 |
| False Positive Rate | 0.0257 | 0.0399 | 0.0303 |
| 95% CI | [0.9846 1.0000] | [0.9589 0.9813] | [0.4897 0.5352] |

Table 3. kNN Performance Locker/SLocker

| Metric | Real vs Benign | Real + Synthetic vs Benign | Trained on Synthetic Tested on Real vs Benign |
|---|---|---|---|
| Accuracy | 0.9788 | 0.9793 | 0.8925 |
| ROC AUC | 0.9919 | 0.9903 | 0.8949 |
| Precision | 0.9844 | 0.9838 | 0.9997 |
| Recall | 0.9730 | 0.9747 | 0.7852 |
| F1 Score | 0.9787 | 0.9792 | 0.8796 |
| False Positive Rate | 0.0154 | 0.0160 | 0.0003 |
| 95% CI | [0.9736 0.9836] | [0.9747 0.9846] | [0.8855 0.8994] |

Table 4. kNN Performance Airpush/StopSMS

## 4.3 Decision Trees

Decision Trees (DTs) are a popular supervised learning approach in malware detection, known for their clear structure and interpretability. They build a hierarchical model by splitting data into branches based on selected features, leading to a final decision such as classifying a record as benign or malicious. This rule-based process is straightforward to follow, which makes DTs particularly useful in security contexts where analysts benefit from understanding why a classification was made (Ferdous *et al.*, 2025). DTs are applied across a variety of malware detection studies, using features that can be static, dynamic or hybrid in nature (Ucci, Aniello and Baldoni, 2019).

Despite these advantages, DTs can be less robust against advanced malware techniques such as obfuscation or polymorphism. This makes them valuable when combined with other models, for example in ensemble methods, to strengthen resilience against evasion strategies (Ferdous *et al.*, 2025). The detailed performance metrics for DT are provided in the tables 5, 6 and 7 below, highlighting key indicators across scenarios. Each table represents results per each individual malware family. The Real vs Benign column represents the baseline scenario where both malware and benign samples were taken from the KronoDroid dataset without synthetic augmentations. The Real + Synthetic vs Benign column represents results where the malware samples from the KronoDroid dataset were enriched with the synthetic data we generated. The Trained-on Synthetic Tested on Real vs Benign column represents the results where the model was trained exclusively using synthetic malware samples we generated and real benign samples and then tested on real malware and benign samples from the KronoDroid dataset. A 95% confidence interval (CI) is a range of values calculated from sample data that, in the long run, would contain the true population parameter in about 95% of repeated samples.

| Metric | Real vs Benign | Real + Synthetic vs Benign | Trained on Synthetic Tested on Real vs Benign |
|---|---|---|---|
| Accuracy | 0.9942 | 0.9911 | 0.6834 |
| ROC AUC | 0.9942 | 0.9910 | 0.6854 |
| Precision | 0.9885 | 0.9882 | 0.9312 |
| Recall | 1.0000 | 0.9940 | 0.3960 |
| F1 Score | 0.9942 | 0.9911 | 0.5557 |
| False Positive Rate | 0.0116 | 0.0119 | 0.0293 |
| 95% CI | [0.9942, 1.0000] | [0.9821 0.9970] | [0.6595 0.7080] |

Table 5. DTs Performance BankBot

| Metric | Real vs Benign | Real + Synthetic vs Benign | Trained on Synthetic Tested on Real vs Benign |
|---|---|---|---|
| Accuracy | 0.9800 | 0.9601 | 0.5672 |
| ROC AUC | 0.9799 | 0.9606 | 0.5822 |
| Precision | 0.9746 | 0.9671 | 0.7870 |
| Recall | 0.9858 | 0.9526 | 0.1842 |
| F1 Score | 0.9802 | 0.9598 | 0.2985 |
| False Positive Rate | 0.0257 | 0.0324 | 0.0498 |
| 95% CI | [0.9686 0.9900] | [0.9464 0.9726] | [0.5439 0.5899] |

Table 6. DTs Performance Locker/SLocker

| Metric | Real vs Benign | Real + Synthetic vs Benign | Trained on Synthetic Tested on Real vs Benign |
|---|---|---|---|
| Accuracy | 0.9781 | 0.9747 | 0.8813 |
| ROC AUC | 0.9829 | 0.9819 | 0.8835 |
| Precision | 0.9775 | 0.9824 | 0.9943 |
| Recall | 0.9788 | 0.9667 | 0.7670 |
| F1 Score | 0.9781 | 0.9745 | 0.8660 |
| False Positive Rate | 0.0225 | 0.0173 | 0.0044 |
| 95% CI | [0.9727 0.9830] | [0.9694 0.9799] | [0.8737 0.8884] |

Table 7. DTs Performance Airpush/StopSMS

**4.4 Logistic Regression**

Logistic Regression (LR) has been widely studied as a machine learning approach for malware detection because of its suitability for large datasets and its efficiency in binary classification tasks. It models the likelihood of an application being malicious or benign and has been applied to both software and hardware-based detection scenarios (Farooq *et al.*, 2022).

From a broader perspective, machine learning techniques like LR are valuable in modern malware detection because traditional methods, such as signature-based detection, are often ineffective against polymorphic and evolving threats. ML enables automated identification of malicious behaviour and reduces detection latency in complex and dynamic environments, which is critical given the growth of malware across PCs, mobile, IoT, and cloud platforms (Ferdous *et al.*, 2025). The detailed performance metrics for kNN are provided in the tables 8, 9 and 10 below, highlighting key indicators across scenarios. Each table represents results per each individual malware family. The Real vs Benign column represents the baseline scenario where both malware and benign samples were taken from the KronoDroid dataset without synthetic augmentations. The Real + Synthetic vs Benign column represents results where the malware samples from the KronoDroid dataset were enriched with the synthetic data we generated. The

Trained-on Synthetic Tested on Real vs Benign column represents the results where the model was trained exclusively using synthetic malware samples we generated and real benign samples and then tested on real malware and benign samples from the Krono-Droid dataset. A 95% confidence interval (CI) is a range of values calculated from sample data that, in the long run, would contain the true population parameter in about 95% of repeated samples.

| Metric | Real vs Benign | Real + Synthetic vs Benign | Trained on Synthetic Tested on Real vs Benign |
|---|---|---|---|
| Accuracy | 1.0000 | 0.9911 | 0.5824 |
| ROC AUC | 1.0000 | 0.9949 | 0.5889 |
| Precision | 1.0000 | 0.9940 | 0.9280 |
| Recall | 1.0000 | 0.9881 | 0.1787 |
| F1 Score | 1.0000 | 0.9910 | 0.2997 |
| False Positive Rate | 0.0000 | 0.0060 | 0.0139 |
| 95% CI | [1.0000 1.0000] | [0.9836 0.9970] | [0.5547 0.6094] |

Table 8. LR Performance BankBot

| Metric | Real vs Benign | Real + Synthetic vs Benign | Trained on Synthetic Tested on Real vs Benign |
|---|---|---|---|
| Accuracy | 0.9757 | 0.9613 | 0.5054 |
| ROC AUC | 0.9906 | 0.9833 | 0.3839 |
| Precision | 0.9718 | 0.9579 | 0.6042 |
| Recall | 0.9801 | 0.9651 | 0.0314 |
| F1 Score | 0.9759 | 0.9615 | 0.0597 |
| False Positive Rate | 0.0286 | 0.0424 | 0.0206 |
| 95% CI | [0.9643 0.9872] | [0.9488 0.9738] | [0.4821 0.5293] |

Table 9. LR Performance Locker/SLocker

| Metric | Real vs Benign | Real + Synthetic vs Benign | Trained on Synthetic Tested on Real vs Benign |
|---|---|---|---|
| Accuracy | 0.9756 | 0.9765 | 0.8495 |
| ROC AUC | 0.9942 | 0.9909 | 0.8631 |
| Precision | 0.9907 | 0.9911 | 0.9906 |
| Recall | 0.9601 | 0.9617 | 0.7058 |
| F1 Score | 0.9752 | 0.9762 | 0.8243 |

| False Positive Rate | 0.0090 | 0.0086 | 0.0067 |
|---|---|---|---|
| 95% CI | [0.9701 0.9810] | [0.9713 0.9815] | [0.8418 0.8575] |

Table 10. LR Performance Airpush/StopSMS

## 4.5 Multilayer Perceptron (MLP)

A Multilayer Perceptron (MLP) is a supervised feedforward artificial neural network and is often described as the base architecture of deep learning models. It is structured as a fully connected network that contains an input layer, an output layer, and one or more hidden layers that perform the main computational tasks. The network generates outputs through activation functions such as ReLU, Tanh, Sigmoid, or Softmax, and is typically trained using backpropagation with optimisation methods including stochastic gradient descent and Adam. Adjusting hyperparameters like the number of hidden layers or neurons can be computationally demanding, but MLPs provide the important advantage of modelling non-linear relationships effectively (Sarker, 2021).

In malware detection, deep neural networks such as MLPs are valuable because their layered structure can learn and extract abstract features from complex, high-dimensional data. This reduces the dependence on manual feature engineering, which is often necessary in traditional machine learning methods. By processing raw input through successive layers, they can identify hidden patterns in behaviours or structures that may indicate malicious activity (Song et al., 2025).

Research shows that neural networks have achieved high levels of accuracy in malware detection across multiple platforms. Detection rates with such models frequently exceed 95%, demonstrating their adaptability and strong performance against evolving and polymorphic malware threats (Ferdous et al., 2025). The detailed performance metrics for MLP are provided in the tables 11, 12 and 13 below, highlighting key indicators across scenarios. Each table represents results per each individual malware family. The Real vs Benign column represents the baseline scenario where both malware and benign samples were taken from the KronoDroid dataset without synthetic augmentations. The Real + Synthetic vs Benign column represents results where the malware samples from the KronoDroid dataset were enriched with the synthetic data we generated. The Trained-on Synthetic Tested on Real vs Benign column represents the results where the model was trained exclusively using synthetic malware samples we generated and real benign samples and then tested on real malware and benign samples from the KronoDroid dataset. A 95% confidence interval (CI) is a range of values calculated from sample data that, in the long run, would contain the true population parameter in about 95% of repeated samples.

| Metric | Real vs Benign | Real + Synthetic vs Benign | Trained on Synthetic Tested on Real vs Benign |
|---|---|---|---|
| Accuracy | 0.9961 | 0.9926 | 0.6410 |
| ROC AUC | 0.9979 | 0.9991 | 0.9108 |
| Precision | 0.9961 | 0.9911 | 0.9791 |
| Recall | 0.9961 | 0.9940 | 0.2881 |
| F1 Score | 0.9961 | 0.9926 | 0.4452 |
| False Positive Rate | 0.0039 | 0.0089 | 0.0062 |
| 95% CI | [0.9903 1.0000] | [0.9851 0.9985] | [0.6133 0.6672] |

Table 11. MLP Performance BankBot

| Metric | Real vs Benign | Real + Synthetic vs Benign | Trained on Synthetic Tested on Real vs Benign |
|---|---|---|---|
| Accuracy | 0.9857 | 0.9751 | 0.5005 |
| ROC AUC | 0.9941 | 0.9869 | 0.3483 |
| Precision | 0.9885 | 0.9635 | 0.5217 |
| Recall | 0.9829 | 0.9875 | 0.0130 |
| F1 Score | 0.9857 | 0.9754 | 0.0254 |
| False Positive Rate | 0.0114 | 0.0374 | 0.0119 |
| 95% CI | [0.9715, 0.9900] | [0.9638 0.9850] | [0.4778 0.5238] |

Table 12. MLP Performance Locker/SLocker

| Metric | Real vs Benign | Real + Synthetic vs Benign | Trained on Synthetic Tested on Real vs Benign |
|---|---|---|---|
| Accuracy | 0.9775 | 0.9775 | 0.8036 |
| ROC AUC | 0.9948 | 0.9939 | 0.8557 |
| Precision | 0.9793 | 0.9813 | 0.9954 |
| Recall | 0.9756 | 0.9735 | 0.6101 |
| F1 Score | 0.9774 | 0.9774 | 0.7565 |
| False Positive Rate | 0.0206 | 0.0185 | 0.0028 |
| 95% CI | [0.9720 0.9826] | [0.9722 0.9824] | [0.7951 0.8128] |

Table 13. MLP Performance Airpush/StopSMS

### 4.6 Random Forest

Random Forest (RF) is an ensemble learning method that constructs multiple decision trees and aggregates their outcomes through majority voting or averaging. This reduces classification errors and helps prevent overfitting, making it more robust than many traditional single classifiers. Each tree is built on random subsets of features, which increases diversity among the trees and strengthens the model's generalization capacity (Farnaaz and Jabbar, 2016).

In intrusion and malware detection contexts, Random Forest has been applied successfully due to its ability to handle high-dimensional data and its resilience against noise. Studies show that it can achieve both high detection rates and low false alarm rates (Farnaaz and Jabbar, 2016).

Ensemble methods like Random Forest offer substantial accuracy and robustness by combining multiple models, complementing deep learning techniques while remaining computationally efficient (Ferdous et al., 2025; Song et al., 2025). Additionally, hybrid approaches have integrated RF with feature selection or deep learning frameworks to improve detection of complex threats, such as Android malware and PowerShell-based attacks (Ferdous *et al.*, 2025; Song *et al.*, 2025). The detailed performance metrics for RF are provided in the tables 14, 15 and 16 below, highlighting key indicators across scenarios. Each table represents results per each individual malware family. The Real vs Benign column represents the baseline scenario where both malware and benign samples were taken from the KronoDroid dataset without synthetic augmentations. The Real + Synthetic vs Benign column represents results where the malware samples from the KronoDroid dataset were enriched with the synthetic data we generated. The Trained-on Synthetic Tested on Real vs Benign column represents the results where the model was trained exclusively using synthetic malware samples we generated and real benign samples and then tested on real malware and benign samples from the KronoDroid dataset. A 95% confidence interval (CI) is a range of values calculated from sample data that, in the long run, would contain the true population parameter in about 95% of repeated samples.

| Metric | Real vs Benign | Real + Synthetic vs Benign | Trained on Synthetic Tested on Real vs Benign |
|---|---|---|---|
| Accuracy | 1.0000 | 0.9985 | 0.6926 |
| ROC AUC | 1.0000 | 0.9998 | 0.9831 |
| Precision | 1.0000 | 1.0000 | 1.0000 |
| Recall | 1.0000 | 0.9970 | 0.3852 |
| F1 Score | 1.0000 | 0.9985 | 0.5562 |
| False Positive Rate | 0.0000 | 0.0000 | 0.0000 |
| 95% CI | [1.0000 1.0000] | [0.9955 1.0000] | [0.6680 0.7173] |

Table 14. RF Performance BankBot

| Metric | Real vs Benign | Real + Synthetic vs Benign | Trained on Synthetic Tested on Real vs Benign |
|---|---|---|---|
| Accuracy | 0.9815 | 0.9913 | 0.5005 |
| ROC AUC | 0.9988 | 0.9988 | 0.7391 |
| Precision | 0.9856 | 1.0000 | 0.5385 |
| Recall | 0.9771 | 0.9825 | 0.0076 |
| F1 Score | 0.9813 | 0.9912 | 0.0150 |
| False Positive Rate | 0.0142 | 0.0000 | 0.0065 |
| 95% CI | [0.9715 0.9900] | [0.9838 0.9963] | [0.4772 0.5233] |

Table15. RF Performance Locker/SLocker

| Metric | Real vs Benign | Real + Synthetic vs Benign | Trained on Synthetic Tested on Real vs Benign |
|---|---|---|---|
| Accuracy | 0.9785 | 0.9815 | 0.8589 |
| ROC AUC | 0.9977 | 0.9974 | 0.9001 |
| Precision | 0.9934 | 0.9949 | 0.9982 |
| Recall | 0.9633 | 0.9679 | 0.7191 |
| F1 Score | 0.9781 | 0.9812 | 0.8360 |
| False Positive Rate | 0.0064 | 0.0049 | 0.0013 |
| 95% CI | [0.9733 0.9830] | [0.9769 0.9861] | [0.8509 0.8668] |

Table 16. RF Performance Airpush/StopSMS

**4.7 Discussion**

The findings highlight both the promise and limitations of using LLM-generated synthetic data for malware detection. Models trained exclusively on real malware and benign data achieved exceptionally high detection performance across BankBot, Locker and Airpush families. For BankBot, multiple classifiers including Logistic Regression and Random Forest achieved perfect performance (test Accuracy = 1.000, ROC AUC = 1.000, precision = 1.000, recall = 1.000, F1 = 1.000). Other classifiers such as kNN and MLP also produced near-perfect results, with accuracies above 0.99 and ROC AUC consistently >0.99. For Locker, performance was similarly strong, however, slightly lower than BankBot. Across classifiers, test accuracy ranged between 0.9715 and 0.9857, with ROC AUC values of 0.986–0.998. For Airpush/StopSMS, real-only models were likewise strong across classifiers (Accuracy ≈ 0.976–0.979; ROC AUC ≈ 0.983–0.998), reinforcing that real data yields near-perfect discrimination. These results demonstrate that when sufficient real malware samples are available, classifiers can achieve near-perfect discrimination between malicious and benign applications.

Augmenting real malware data with LLM-generated synthetic samples produced results that were also strong, but generally slightly lower than using real data alone. For BankBot, test accuracy for the best models ranged from 0.9911 to 0.9985, with ROC AUC between 0.9910 and 0.9998. For Locker, accuracies ranged from 0.9601 to 0.9751, with ROC AUC between 0.9606 and 0.9869. For Airpush/StopSMS the performance remained excellent and very close to real-only (Accuracy ≈ 0.975–0.982; ROC AUC ≈ 0.982–0.997), again slightly below the real-only baseline. While these results remain strong, they consistently fall short of the near-perfect performance observed with real-only training. Importantly, the false positive rate remained very low (typically <0.04), indicating that synthetic augmentation does not compromise specificity. However, the slight decline in accuracy and recall suggests that the synthetic data introduces some noise or distributional mismatch relative to the real malware, thereby diluting the effectiveness of the models compared to training exclusively on real samples.

When models were trained exclusively on synthetic data and then evaluated on real malware, performance degraded substantially. For BankBot, classifiers achieved moderate test accuracy (≈0.64–0.74) and ROC AUC values up to 0.98, but recall was consistently poor (≈0.29–0.49), indicating that the models missed a large proportion of real malware instances despite maintaining high precision (≈0.93–1.00). For Locker/SLocker, performance was near random (Accuracy ≈ 0.50–0.57; ROC AUC often < 0.60) with extremely low recall, rendering these models impractical. In contrast, Airpush/StopSMS synthetic-only models generalised substantially better (Accuracy ≈ 0.80–0.89; ROC AUC ≈ 0.86–0.90; Recall ≈ 0.61–0.79), while maintaining very high precision (≈ 0.99). This improvement coincides with the larger family size and deeper LLM finetuning (150 samples, 3 epochs), indicating that synthetic-only generalisation is sensitive to family characteristics and finetuning regime.

Real-only training remains the benchmark, with near-perfect results across all three families. Real with synthetic augmentation preserves high performance but is consistently, slightly below real-only. Synthetic-only training is family and method-dependent: weak for Locker/SLocker, moderate for BankBot, and stronger for Airpush/StopSMS under a larger/deeper finetuning setup. These results suggest that improving synthetic fidelity via more finetuning data/epochs for larger families can materially narrow the synthetic to real generalisation gap.

## 5 Conclusions and Future Work

This research set out to evaluate the feasibility of using LLM-generated synthetic malware to support Android threat detection. By positioning real-only detection accuracy as a benchmark, the study contextualised the effectiveness of both augmentation and synthetic-only training. The results show that while LLMs can generate structurally consistent malware records and provide meaningful augmentation, they do not yet achieve the realism or diversity needed to serve as a standalone data source. Notably, for Airpush/StopSMS, a larger family that we fine-tuned with 150-sample and 3 epochs, synthetic-only models achieved substantially higher generalisation than for the other families, indicating that synthetic utility improves with family size and finetuning depth.

The contribution of this work lies less in raw performance metrics and more in the insights it provides for practice. First, it demonstrates that synthetic augmentation can be applied without undermining specificity, offering a practical way to enrich scarce datasets. Second, it reveals the fragility of synthetic-only training, underscoring the gap between generated records and operationally valid malware. Third, it highlights methodological considerations such as prompt design, post-processing, and validation which are crucial when applying LLMs to structured cybersecurity data. These findings establish a foundation for future exploration of synthetic data as both a supplement and a research tool in security analytics.

Future work should systematically ablate finetuning variables: number of samples, epochs, and prompt schema across multiple families to quantify their effect on synthetic fidelity and synthetic to real generalisation. Larger finetuning corpora and family-diverse prompts may reduce distributional drift and improve recall without inflating false positives. Providing larger training sets for fine-tuning and increasing the number of epochs could help improve the quality of data and in turn increase generalisation. Statistical validation methods such as Kolmogorov–Smirnov tests, clustering, or feature-wise correlation analysis could be applied to quantify fidelity. Exploring other LLMs, especially ones that can handle tabular data well, could enhance structural consistency. Integration with adversarial training pipelines and real-time sandbox simulations could offer another avenue to generate context-rich synthetic behaviours.

In conclusion, the findings of this study suggest that LLM-generated malware can be a useful augmentation tool but should not yet be relied upon as a primary data source. Augmenting real datasets with carefully validated synthetic samples can help address data scarcity, especially for rare families or behaviours, without compromising detection specificity. However, deploying classifiers trained exclusively on synthetic data would be premature, as current LLM outputs do not generalise reliably to real-world threats. In practice, synthetic malware is best applied as a complement to, rather than a substitute for, real data-supporting tasks such as adversarial training, red-teaming, and benchmarking in environments where access to sensitive datasets is limited.